\documentstyle[aps,prl,multicol,fancyheadings,epsfig,amsbsy,amssymb,amstex]{revtex}

\def\bbbc{{\mathchoice {\setbox0=\hbox{$\displaystyle\rm C$}\hbox{\hbox
to0pt{\kern0.4\wd0\vrule height0.9\ht0\hss}\box0}}
{\setbox0=\hbox{$\textstyle\rm C$}\hbox{\hbox
to0pt{\kern0.4\wd0\vrule height0.9\ht0\hss}\box0}}
{\setbox0=\hbox{$\scriptstyle\rm C$}\hbox{\hbox
to0pt{\kern0.4\wd0\vrule height0.9\ht0\hss}\box0}}
{\setbox0=\hbox{$\scriptscriptstyle\rm C$}\hbox{\hbox
to0pt{\kern0.4\wd0\vrule height0.9\ht0\hss}\box0}}}}
\newcommand{\beqa}{\begin{eqnarray}}
\newcommand{\eeqa}{\end{eqnarray}}

\newcommand{\bS}{{\bf S}}
\newcommand{\bB}{{\bf B}}
\newcommand{\bk}{{\bf k}}

\newcommand{\br}{{\bf r}}

\newcommand{\bsig}{\mbox{\boldmath{$\sigma$}}}

\begin{document}

\title{Theory of single spin detection with STM}
\author{A.V. Balatsky and I. Martin}
\address{Theoretical Division, Los Alamos National Laboratory, Los
Alamos, NM 87545}

\date{Printed \today }

\maketitle

\begin{abstract}
We propose a mechanism for detection of a single spin center on a
non-magnetic substrate.  In the detection scheme, the STM tunnel
current is correlated with the spin orientation.  In the presence
of magnetic field, the spin precesses and the tunnel current is
modulated at the Larmor frequency. The mechanism relies on the
effective spin-orbit interaction between the injected unpolarized
STM current and the local spin center, which leads to the nodal
structure of the spatial signal profile.  Based on the proposed
mechanism, the strongest spin-related signal can be expected for
the systems with large spin-orbit coupling and low carrier
concentration.

\end{abstract}

\pacs{Pacs Numbers:
rosengren paper}
\vspace*{-0.4cm}
\begin{multicols}{2}

\columnseprule 0pt

\narrowtext \vspace*{-0.5cm}

There is no fundamental principle that precludes the single spin measurement.
Possibility of a single spin observation is therefore a question of spatial and
temporal resolution. The standard electron spin measurement technique --
electron spin resonance -- is limited to a macroscopic number of electron spins
-- $10^{10}$ or more\cite{ESR}; the state-of-the-art magnetic resonance force
microscopy has recently achieved the resolution of about 100 fully polarized
electron spins\cite{Sidels}. It has already been shown that optically induced
ESR of a single spin is  possible \cite{Kohler}. There are proposals for the
spin detection using single electron transistor and spin-polarized current
\cite{Kane2,Loss}. In this letter we propose the theoretical basis
for the new spin-detection technique -- electron spin precession
scanning tunneling microscopy (ESP-STM) -- capable of single spin
detection.

The applications of single spin detection and manipulation range from the study
of strongly  correlated systems, to nanotechnology, to quantum information
processing. In the strongly  correlated systems, the ability to detect single
spin will allow one to investigate the  magnetism on the nano scale by
detecting the changes in the spin behavior while entering magnetically ordered
states \cite{Wiesendanger}. One could also explore temperature evolution of the
magnetic properties of a single paramagnetic atom in the Kondo
regime\cite{Manoharan}. Magnetic properties of spin centers in superconductors
is another area where single spin plays an important role by generating
intragap impurity states \cite{Salkola1,Yazdani}.   In nanotechnology, spins
can be used as elementary information storage units. In the realm of quantum
computing, several architecture  proposals rely on the ability to manipulate
and detect single spins\cite{Kane,q-dots}. Here we argue that STM offers a
powerful technique to detect a {\em single} spin.

\begin{figure}[htbp]
\begin{center}
\includegraphics[width = 3.0 in]{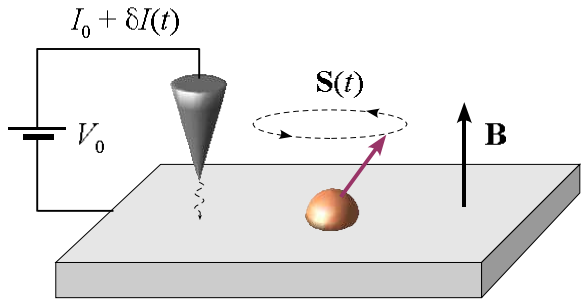} \vspace{0.5cm}
\caption{Experimental setup for the electron spin precession STM.
In the applied magnetic field ${\bf B}$, spin of the magnetic atom
(e.g. Gd, shown in gold) is precessing around the field direction.
When precisely positioned next to the spin site (within a few
\AA), the STM tip can pick up an ac modulation of the tunnel
current. }
\label{fig:setup}
\end{center}
\end{figure}

The electron spin can be detected through its coupling to the
magnetic field, which can either freeze the spin along  the field,
or make it precess around the field direction.  To detect the
single-spin signal, the superb spatial and temporal resolutions
are required. The necessary level of sensitivity can be achieved
in the Scanning Tunneling Microscope (STM).  It is well known that
spatial resolution of the STM is in the sub-\AA ngstr\"om range.
Therefore, the technique naturally lends itself for the detection
of the single spin.  In the STM setup, the spin precession in an
external magnetic field can be detected through an ac modulation
of the tunnel current. In this purely dc configuration, the ac
tunnel current at the precession, or {\em Larmor}, frequency can
be generated due to the effective coupling of the precessing
localized spin to the tunneling electrons.  Hence we argue that
STM can operate as {\em Larmor frequency generator}. The
experimental setup we consider is given in Fig.~\ref{fig:setup}.

Our work is motivated by experiments of Manassen {\it et al.}
\cite{Manas1}.  In these experiments, STM was used to measure the
tunneling current while scanning the surface of Si in the vicinity
of a local spin impurity (Fe cluster) or imperfection (oxygen
vacancy in  Si-O).   A small signal in the current power spectrum
at the Larmor frequency was detected. The ac signal was spatially
localized at the distances on the order of $5-10$ \AA\ from the
spin site. The extreme localization of the signal and the linear
scaling of its frequency with the magnetic field prompted Manassen
{\it et al.} to attribute the detected ac signal to the Larmor
precession of a single spin site.

Rather than going into further details of these experiments, we
pose here a general question:  {\em Under what conditions it is
possible to detect the single precessing spin with STM}?
Surprisingly, we find that a number of mechanisms can generate a
detectable ac current.   The interaction that provides the
coupling of the precessing spin to the tunneling current is the
spin-orbit (SO)  interaction.  Based on our analysis, we argue
that single spin detection is within the reach  of current STM
technology and can be realized in a variety of non-magnetic
materials with isolated paramagnetic centers.

Consider a localized magnetic site with spin  $\bS, S = 1/2$.  In
the presence of magnetic field, $\bB$, the spin-up ($E_\uparrow$)
and spin-down ($E_\downarrow$)  energy levels are Zeeman split. At
zero temperature, only the lowest energy  state (parallel to
magnetic field) is occupied; at a finite temperature, or due  to
an external excitation, the spin may be driven into the mixed
state  characterized by the wavefunction, $|\psi\rangle =
\alpha|\uparrow\rangle +  \beta|\downarrow\rangle$, with the time
evolution $\alpha =  |\alpha|\exp(-iE_{\uparrow}t)$ and $\beta =
|\beta|\exp(-iE_{\downarrow}t +  i\phi(t))$.  The drifting phase
$\phi(t)$ determines the spin  coherence time  $\tau_\phi$ and is
related to the ESR\cite{comment1} spin relaxation time  $T_2$.  In
such a state, the spin expectation value, $\langle
\psi(t)|\bS|\psi(t)\rangle/\langle\psi(t)|\psi(t)\rangle$, will
precesses around the direction of the magnetic field at the Larmor
frequency, \beqa \hbar \omega_L = E_{\uparrow} -E_{\downarrow} = g
\mu_B B, \eeqa where $g$ is the gyromagnetic ratio and $\mu_B$ is
the Bohr magneton. In magnetic field of 100 Gauss, this frequency
for a free electrons is 280 MHz. We assume hereafter that in the
course of the ESP-STM experiment the spin is $not$ in the
groundstate either due to a thermal fluctuation or an external
excitation by the tunneling current.  We also assume that
$\tau_\phi$ is sufficiently long for the spin precession to be
well defined, $\omega_L\tau_\phi \gg 1$.  The validity of these assumptions is
verified at the end of the paper.

We present now a mechanism for the ac STM current generation with no ac input
based on the time-dependent modification of the tunneling density of states
induced near the precessing spin in the presence of applied current.  On the
time scale of all conduction-electron processes, the spin precession is very
slow.  This follows from the comparison of the energy associated with the rf
spin precession, $\hbar\omega_L \sim 10^{-6}$ eV, and the typical electronic
energy scale in metals or semiconductors, which is on the order of 1 eV. Hence,
for every ``instantaneous'' spin orientation the electronic problem can be
solved as if the local moment is static.  This is similar to the
Born-Oppenheimer approximation for atoms where atomic nuclear dynamics is much
slower than the electronic motion.  We describe the system of electrons
interacting with the local impurity spin by the Hamiltonian
\beqa\label{eq:H}
H = H_0 + J\bS\cdot{\bsig}(0),
\eeqa
where $J$ is the strength of the exchange interaction between the
local spin and  conduction electron spin density, ${\bsig}(0)=
\bsig_{\alpha\beta}c^\dagger_\alpha(0)c_\beta(0)$, on the impurity
site.  To provide the coupling of the spin orientation to the
orbital degrees of freedom of electrons, the non-interacting part
of the Hamiltonian, $H_0$, should include spin-orbit interaction.
We will focus here on the two-dimensional band of states on the
material surface. From the symmetry  considerations, the energy of
the surface states contains spin-orbit part that is linear both in
the electron spin, $\bsig$, and momentum, $\bk$, \cite{Rashba}
\beqa
\label{eq:Ek}
\epsilon_{\alpha\beta}(\bk) = \frac{k^2}{2m^*} + g_{\rm
SO}[\bk\times{\bsig}_{\alpha\beta}]_{\hat{n}}.
\eeqa
Here, $m^*$ is the band mass of the surface states and $g_{\rm
SO}$ is the SO coupling strength.  The cross-product of the spin
and momentum is projected onto  the normal to the surface,
${\hat{n}}$.

The problem specified by equations (\ref{eq:H}) and (\ref{eq:Ek})
can be solved  for each instantaneous  value of the precessing
spin $\bS(t)$.  Specifically, we are looking for the correction to
the conduction electron density of states due to the local spin.
The Green function matrix for the conduction electrons is \beqa
\hat{G}_0(\bk, \omega) = [\omega - \hat{\epsilon}(\bk)]^{-1}.
\eeqa In the presence of the current flow ${\bf j}$ in the system,
the equilibrium momentum distribution is shifted by an amount
proportional to the current, $\bk_0 = {\bf j}m^*/ne$, where $n$ is
the carrier density and $e$ is electron charge.  This shift can be
approximately introduced by modifying the electronic Green
functions\cite{lipa}, \beqa \hat{G}(\bk,\omega) = \left[\omega -
\frac{(\bk-\bk_0)^2}{2m^*} - g_{\rm
SO}[\bk\times{\hat{\bsig}}]\right]^{-1}. \eeqa Here, only  the
kinetic energy momentum is shifted since  SO term is explicitly
non-invariant under shift ($\bk \rightarrow {\bf k-k_0}$) as it
measures the SO effects  due to lattice.  The correction to the
density of states near the Fermi surface caused by the impurity
can be then calculated straightforwardly in the first order in the
scattering potential, \beqa \delta N(\br) = \frac{1}{\pi} {\rm Im}
\sum_{\alpha\beta\gamma}{G_{\alpha\beta}(\br)
J\bS\cdot{\bsig}_{\beta\gamma} G_{\gamma\alpha}(-\br)} \eeqa
Expanding in the strength of the SO interaction relative to the
Fermi energy, the $\bS$-dependent contribution to the density of
the surface states obtains in the first order in the exchange
coupling and the strength of SO interaction, \beqa\label{eq:dN}
\frac{\delta N}{N} =  g_{\rm SO} J \frac{dN}{dE} J_0^2(k_F r)
[\bk_0\times\bS]_{\hat{n}}. \eeqa This correction depends on the
distance from the spin center, $r$, through the Bessel function of
the first kind, $J_0(x)$.  This result also follows from
considering Friedel spin oscillation of electrons coupled to the
tunneling DOS in the presence of the SO term\cite{Friedel}. It is
time-dependent in the presence of magnetic field since the
projection of $\bS$ oscillates at the Larmor
frequency\cite{zeeman}.  This expression is consistent with
general properties of DOS.  Since $N({\bf r},t)$ is a scalar, it
should be invariant under time reversal, whereas $\bf S$ is odd
under time reversal.  Hence $\delta N({\bf r},t)$ can depend only
on product of the spin vector with some other vector that is {\em
odd} under time reversal or on time derivative of spin vector.
Possible combinations include $\delta N \sim [{\bf I} \times {\bf
S}]_{\hat{n}}$ (Eq.~{\ref{eq:dN}}) and $\delta N \sim [\partial_t
{\bf S}(t) \times {\bf r}]_{\hat{n}}$. Indeed, we have found
mechanism for the latter contribution as well; however, for the
experimental conditions of Refs.~\cite{Manas1} the time-dependent
DOS mechanism described here dominates.

The current in the STM tunnel junction is proportional to the
single electron density of states in the substrate. Therefore, the
ac current contribution relative to the base dc STM current, $I$,
can be estimated as
\beqa\label{eq:dI}
\frac{\delta I(t)}{I} \sim \frac{\delta N(t)}{N}.
\eeqa
Notice, that the magnitude of the effect is proportional to the current in the
system via $\bk_0$.  Experimentally, the necessary bias current can either be
provided externally (by extra leads) or can be injected by STM itself.  In the
latter case, the direction of the current will vary in space and, in the
absence of anisotropy, will flow directly away from the tip-sample junction.
From the symmetry considerations, in Eq.~(\ref{eq:dN}), $\bk_0$ should be
oriented from the tip towards the spin center. The magnitude of the equilibrium
shift is the 2D case considered here is $k_0\sim m^*I/nr$, where $n$ is the 2D
density of the surface carriers. Since $\bk_0$ itself is proportional to
injected current, the $\bS$-dependent correction to the current is proportional
to $I^2$.  For direct tunneling into the spin center, $\bk_0$ is parallel to
the tunneling axis and hence there is no $\bS$-dependent contribution to the
current. This {\em nodal} structure is a consequence of the problem symmetry.
The magnitude of the time-dependent density of states is independent of the
magnetic field orientation.

\begin{figure}[htbp]
\begin{center}
\includegraphics[width = 3.0 in]{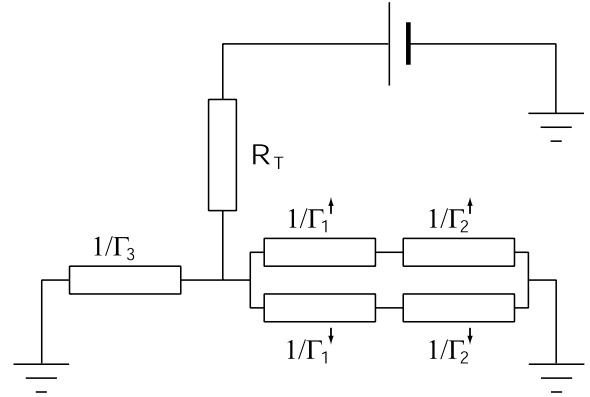} \vspace{0.5cm}
\caption{Electrical circuit representing ESP-STM experiment.  The
current is injected through the tunnel junction of resistance
$R_T$.  Inside the material the current can either flow through
the impurity site (right, ``impurity channel'') or directly to the
leads (left). In the impurity channel, $\bS$-dependence of the
total current obtains through spin-current polarization due to
spin-orbit coupling (conductance $\Gamma_1^\uparrow \ne
\Gamma_1^\downarrow$) and spin-dependent transmission through the
impurity site ($\Gamma_2^\uparrow \ne \Gamma_2^\downarrow$).  The
resulting $\bS$-dependent contribution to the total circuit
current, Eq.~(\ref{eq:dIc}), is in agreement with the more
sophisticated calculation, Eq.~(\ref{eq:dN}). }
\label{fig:circ}
\end{center}
\end{figure}

The intuitive understanding of this result can be obtained using an analogue of
the spin-dependent B\"{u}ttiker-Landauer formalizm\cite{but,rouk}.  The
electrical circuit that corresponds to the STM experiment is shown in Fig.
\ref{fig:circ}.  When current is injected through the tunnel junction with
resistance
$R_T$, it can either flow to the leads directly (left branch
with conductance $\Gamma_3$) or through the channel affected by
the spin center (right branch, $\Gamma^\sigma_1$ and
$\Gamma_2^\sigma$).  We assume here that there are no spin flips.
From geometric considerations, the closer tip is laterally to the spin center,
the larger is the ratio, $\Gamma_{1(2)}/\Gamma_3$. From the form of SO
interaction, Eq.~(\ref{eq:Ek}), the relevant spin $\sigma$ quantization axis
for electrons flowing from the tunneling point to the impurity is perpendicular
both to the surface normal, $\hat{n}$, and the vector connecting these two
points, $\bk_0$.  The essence of the right branch of the circuit is that in the
presence of SO interaction the current becomes spin polarized,
$\Gamma_1^\uparrow \ne \Gamma_1^\downarrow$, and the transmission of the local
magnetic center is spin sensitive,
$\Gamma_2^\uparrow \ne \Gamma_2^\downarrow$.  Approximately,
\beqa
\Gamma_1^\sigma &=& \Gamma(1+g_{SO}k_0\sigma/E_F),\\
\Gamma_2^\sigma &=& \Gamma(1+JS\sigma/E_F),
\eeqa
where $E_F$ is the Fermi energy of the substrate, and S is the projection of
spin $\bS$ onto the $\sigma$ quantization axis (above). Similar to the Green
function treatment presented above, the expression for $\Gamma_1^\sigma$ goes
beyond linear response since it depends on the current through $k_0$. The
correction to the current due to the spin-selectivity of the circuit is
\beqa\label{eq:dIc}
\frac{\delta I}{I} = \frac{R_M^2\Gamma}{R_M + R_T} \frac{g_{SO}k_0
J}{E_F^2}S,
\eeqa
where $R_M = 1/(\Gamma + \Gamma_3)$.  This expression is the same as the
previously obtained result, Eqs.~(\ref{eq:dN}) and (\ref{eq:dI}).

From Eq.~(\ref{eq:dN}) it is clear that the largest ESP-STM effect can be
obtained in the systems with large spin orbit interaction and strong
particle-hole asymmetry in the density of states. Small density of surface
carriers, $n$, helps to drive the carrier system out of equilibrium, which is
crucial for the ESP-STM effect. All these parameters can be tuned by the choice
of the substrate material, as well as the type of the spin center, to achieve
large response. For model parameters,
$g_{\rm SO} k_F \sim 0.01$~eV, $J\sim 1$~eV, $dN/dE\sim 1$~eV$^{-2}$, and Fermi
surface displacement $k_0/k_F\sim 0.1$, the effect in the density of states is
$\delta N/N \sim 10^{-3}$, which should be experimentally detectable.

In the above discussion we assumed that the spin dephasing time, $\tau_\phi$,
is long compared to the Larmor period, which implies weak coupling to
environment.  However, even under ideal circumstances, the measurement will
induce spin dephasing.  This
$backaction$ is an unavoidable side effect of measurement.  We will now
estimate the backaction of the tunneling electrons on the precessing spin. By
applying a canonical transformation, the original Hamiltonian,
Eqs.~(\ref{eq:H}) and (\ref{eq:Ek}), can be mapped to a model of two reservoirs
($L$ and $R$) of spinless electrons coupled through a spin-dependent tunneling
matrix element\cite{CT},
$$H_{\rm eff} = H_L + H_R + h_z S_z + \sum_{l\in L,\ r\in R}c^{\dag}_{l}[T_0 + T_1 S_x]c_r. $$
Here $T_0$ is the tunneling amplitude that doesn't involve local spin and
$T_1 \simeq {g_{so} k_0 J }(dN/dE)\, T_0$ is the spin-dependent part of the tunneling.  The
$x$ axis is chosen along $\hat{n}\times {\bf k}_0$ direction.
For spin-1/2 quantum systems, this model has been extensively
studied\cite{gurvitz,KA}.  Two important parameters in the model are the
electron tunneling rate, $1/\tau_e = \pi T_0^2 N_L N_R V$, and the
measurement-induced spin-dephasing rate, ${1}/{\tau_s} = \pi T_1^2 N_L N_R V$.
It has been demonstrated\cite{KA} that under the {\em weak measurement}
conditions, $\hbar/\tau_s < h_z$, the stationary power spectrum of the tunnel
current has a peak at the Larmor frequency,
\begin{equation}\label{eq:dIdI}
\frac{\langle I^2_{\omega} \rangle} {2e I_0} =
\frac{4/\tau_s^2}{(\omega - \omega_L)^2 + 1/\tau_s^2}
\end{equation}
For weak measurement, increasing applied voltage does not change the peak value
relative to the shot noise level; however it increases the weight under the
resonance peak.  Under the stationary conditions, the coupling to the current
fluctuations also provides the excitation for the spin.  For the same
parameters as above we find spin dephasing time $1/\tau_s
\sim 10^4 \ldots 10^5\ {\rm Hz} \ll \omega_L$.
Experimentally, the observed linewidth is narrow,
$\sim 500\ {\rm KHz} \ll \omega_L$ \cite{Manas1}, which is consistent with our effective model.
Any extrinsic (unrelated to measurement) dephasing mechanisms will decrease the
signal to shot noise ratio.

The applicability of the  presented ESP-STM mechanism is not limited to
semiconductors.  Other systems with stronger spin-orbit interaction, such  as
Au or Cu surfaces\cite{Manoharan} with local magnetic defects may be expected
to show a similar effect. Yet another route is to use heavy  magnetic atoms,
such as Eu or Gd as the local spin impurities with large  intrinsic SO
interactions. In this case, a result similar to Eq.~\ref{eq:dN} obtains, with
the strength of the effect proportional to the spin-orbit coupling  on the
impurity site.  In general, any form of SO interaction is capable of coupling
the precessing local spin to the itinerant electron density of  states, which
can be measured with STM.

The spin-current coupling mechanism that we described applies to the static
(dc) case as well.  The dc case can be realized by applying a strong polarizing
magnetic field and detecting the variation of the tunneling current as a
function of the magnetic field orientation.  This, however, is only possible at
low temperatures, due to the limited strength of available magnetic fields.
The dc measurement is further complicated by the low-frequency (1/f) noise.

In conclusion, we have proposed a mechanism for the novel technique for the
single spin detection with STM.  It is based on the ac modulation of the
tunneling current caused by the effective spin-orbit coupling to the precessing
local spin.  From the analysis of the coupling mechanism we identified classes
of materials and experimental conditions for which significant single spin
signal can be expected.

We are grateful to G. Boebinger, F. Bronold, G. Brown, J.C. Davis,
D. Eigler, C. Hammel, M. Hawley, Y. Manassen, A. Migliory, D.
Mozyrsky, Y. Nazarov,  D. Pines and A. Yazdani for useful
discussions.  We are grateful to D. Mozyrsky for pointing out
unversal features of the current spectral density. This work was
supported by the U.S. DoE.


\end{multicols}


\begin{thebibliography}{99}
\bibitem{zeeman}
The effect of the magnetic field on electrons in the conventional metallic STM
tip can be neglected due to the vanishingly small magnitude of the Zeeman
energy relative to the Fermi energy;  the effect on the sample surface carriers
can be neglected as well if the external magnetic field is relatively weak,
such that $\mu_B B < g_{SO}k_F$.  This condition is satisfied in the systms
with strong spin-orbit coupling that we are interested in.
\end{thebibliography}
\end{document}